\documentclass{article}
\usepackage{graphicx} 
\usepackage[a4paper, total={18cm, 25cm}]{geometry}
\usepackage{longtable}
\usepackage{rotating}
\usepackage{csquotes}
\usepackage[style=ieee,citestyle=numeric-comp, backend=biber]{biblatex}
\usepackage[version=4]{mhchem} 
\usepackage{hyperref}
\usepackage{hyperxmp}
\usepackage{booktabs}
\usepackage{siunitx}
\ifdefined\unit\else
 \ifdefined\NewCommandCopy
    \NewCommandCopy\unit\si
 \else
    \NewDocumentCommand\unit{O{}m}{\si[#1]{#2}}
 \fi
\fi

\usepackage{multicol}  
\usepackage{enumitem}  
\usepackage{adjustbox}  
\usepackage{colortbl}  
\newcolumntype{\darkgreencolumn}{>{\columncolor[HTML]{7DFA7D}} p{0.22\textwidth}}
\newcolumntype{\lightgreencolumn}{>{\columncolor[HTML]{BBFA7D}} p{0.22\textwidth}}
\newcolumntype{\yellowcolumn}{>{\columncolor[HTML]{FAFA7D}} p{0.22\textwidth}}
\newcolumntype{\orangecolumn}{>{\columncolor[HTML]{FABC7D}} p{0.22\textwidth}}
\newcolumntype{\redcolumn}{>{\columncolor[HTML]{FA7D7D}} p{0.22\textwidth}}
\usepackage{pifont}
\newcommand{\tick}{\ding{51}}%
\newcommand{\cross}{\ding{55}}%

\title{Sustainability Assessment of Future Accelerators}
\author{Input to the European Strategy process \\by the LDG Working Group on Sustainability  \vspace{1cm} \\edited by C.\,Bloise\footnote{caterina.bloisel@lnf.infn.it}, M.\,Titov\footnote{maxim.titov@cea.fr} \\ 
with contributions from  E.\, Cennini, J.\,Gutleber, W.\,Kaabi, A.\,Klumpp,\\ P.\,Koppenburg, Y.\,Li, B.\,List, R.\,Losito, B.\,Mandelli, E.A.\,Nanni, N.\,Neufeld,\\ T.\,Schoerner-Sadenius, B.\,Shepherd, V.\,Shiltsev, S.\,Stapnes, L.\,Ulrici, H.\,Wakeling} 

\date{30 March 2025}

\begin{document}
\maketitle

\begin{abstract}
  
The Large Particle Physics Laboratory Directors Group (LDG) established the Working Group on the Sustainability Assessment of Future Accelerators in 2024 with the mandate to develop guidelines and a list of key parameters for the assessment of the sustainability of future accelerators in particle physics. While focused on accelerator projects, much of the work will also be relevant to other current and future Research Infrastructures.

The development and continuous update  of such a framework aim to enable a coherent communication amongst scientists
and  adequately convey the information to a broader set of stakeholders. 
 This document outlines the major findings and recommendations from the LDG Sustainability WG report - a summary of current best practices recommended to be adopted by new Research Infrastructures. The full report will be available in June 2025 at: https://ldg.web.cern.ch/working-groups/sustainability-assessment-of-accelerators. 

Not all of sustainability topics are addressed at the same level. The assessment process is complex, largely under development and a homogeneous evaluation of all the aspects deserves a strategy to be pursued over time.

\end{abstract}

\maketitle

\clearpage
\sffamily

\section{Executive Summary}

Establishing and maintaining a reference framework for the sustainability assessment of future accelerators is intended to provide a consistent understanding of the evaluation process to promote transparent, open discussions within research communities and to adequately convey the information to a broader set of stakeholders. 

Large research infrastructures (RIs) featuring particle accelerators designed to explore fundamental interactions at the energy frontier face the challenge of meeting demanding performance requirements essential to their scientific mission, while also ensuring sustainable construction and operation. 
A long-term vision and approach is needed for all aspects of the project, including sustainability assessment. 
RIs are typically designed and operated for several decades, so assessment 
requires continuous efforts to maintain and update guidelines and reporting values.

Sustainability assessment is complex, in that criteria must be tuned to the maturity of the projects and 
developed separately for scientists, for policy and decision takers, and for society. 
International organizations and fora are working to establish the goals and scope of sustainable development
projects in different sectors relevant to accelerator-based RIs. 
Economists, climate change experts, and environmental engineers have contributed a range of work providing guidelines within their respective fields, which inform international, European, and national legislation in setting standards and obligations aimed at advancing towards the objectives of the Paris Agreement. 

As challenging global technological enterprises, accelerators-based RIs generate socio-economic impacts beyond scientific knowledge, across several of the 17 Sustainable Development Goals (SDGs) in the 2030 Agenda for Sustainable Development adopted by United Nations Member States in 2015. These include: good health and well-being (3); high-quality education (4); industry, innovation and infrastructure (9); climate action (13); peace, justice and strong institutions (16); partnerships for the goals (17). Understanding of these impacts is necessary for public authorities and credit institutions to make informed investment decisions. Comprehensive sustainability assessments based on quantitative Cost-Benefit Analysis (CBA) have been carried out and are required by several countries for public investments, including science projects. 
At the level of the European Strategy Forum for Research Infrastructures (ESFRI), “socio-economic impact has become one of the important considerations in the roadmapping process that identifies European investment priorities in Research Infrastructures” and has also increasingly entered the discussions on funding priorities at national/regional level. 

The goal of designing sustainable accelerator-based RIs is intended to align with the broader objectives of the Sustainable Development Goals (SDGs).
Public authorities have developed guidelines for presenting infrastructure projects that are becoming increasingly demanding concerning environment protection. Following this direction, ESFRI underlines that the “ESFRI Roadmap 2026 will put significant focus on the financial sustainability of the RI ecosystem, while introducing the new dimension of ‘environmental considerations’ to foster environmental sustainability”.

Life Cycle Assessment (LCA) is a methodological approach to quantify environmental impact defined by the
ISO14040 standard suite and is the subject of section \ref{sec:lca}. The goals for conducting an LCA range from understanding the most critical elements in the design at an early project stage, to providing input for technological choices, to fulfilling legal requirements for project submission. Different goals affect the methodology and metrics adopted, including the scope
and functional units, life cycle stages considered, and impact categories studied. 
Performing an LCA for the major components and contributions at every stage of design (proposal, CDR, TDR) is recommended. 
LCAs carried out at different levels of maturity of the project serve distinct scopes. 
At an early stage of the infrastructure project, the assessment is limited, but has the benefit of embedding sustainability into the design process.
The goals and scope also depend on the target audience, aimed at design studies when addressed to scientists, at project review when directed to decision makers, and to communicate global performance when it comes to society. 
Coordinated efforts from all accelerator-based laboratories to establish a centralized database of materials and components in relation to accelerators, including, when possible, comprehensive LCAs from mining to fabrication, are strategic to establish a framework for project accountability.
Assumptions/simplifications adopted, and sources of uncertainty within the LCAs, such as accuracy of input data and accuracy of extrapolations with time, deserve proper consideration and reporting to clarify the boundaries of the assessment.

The impact of emissions on climate change, quantified as Global Warming Potential (GWP), is by far the most considered in public debate and legislation. GWP is present in all series of midpoint categories used for LCAs that include amongst others, emissions of particulates, use of freshwater, land use and transformation, and use of natural resources. On the \emph{construction phase}, engineering civil works are a major source of emissions, deserving particular attention in assessing the sustainability of the project, followed by materials and manufacturing needed for RF systems and magnets. 
For the \emph{operation phase}, electricity consumption (RF, magnets, cooling, services) is the most relevant indirect contribution to GWP, keeping in mind that CO$_2$ footprint of electricity varies from one region to another and with time. 
The \emph{decommissioning phase} requires analysis of recycling and disposal of used components, radioactive waste, and land reuse. 
Mitigation and compensation measures (the subject of section \ref{sec:mitigation}), include the development of energy management plans (EnMS), the introduction of low-environmental footprint materials and procedures for engineering works, the responsible
procurement of products and electricity, the implementation of systems for heat recovery and supply, the investment in R\&D of green accelerator technologies, and nature-based interventions.

The LDG WG on the Sustainability Assessment of Accelerators comprises researchers working on new accelerator-based RIs, members of sustainability panels at CERN and other laboratories worldwide and representatives of initiatives with a focus on the sustainability of accelerators, funded by the European Commission (iFAST, iSAS, EAJADE). The full report will be available in June 2025 at: https://ldg.web.cern.ch/working-groups/sustainability-assessment-of-accelerators.

\section{Building Strategic Accountability}\label{sec:account}

Sustainability assessment is based on the analysis of main three pillars: society, economy, and environment.
To achieve sustainable science, the project appraisal integrates financial and socio-economic aspects.
Socio-economic aspects comprise social benefits (positive externalities) as well as negative effects such as climate impacts, consumption of natural resources such as soil, land and water, impacts on the quality of air and water, on public health and numerous other externalities.
Future accelerators are international initiatives requiring RIs of a size and cost deliverable only through international collaboration and   represent significant long-term investments for nations. 
Starting with their conceptual design, an appraisal process is typically requested by funding agencies, public authorities, 
and credit institutions.
Periodical monitoring and tracking of social, environmental and economic performance against initial evaluations can trigger a continuous process of improvement.
A set of legal frameworks and appraisal guidelines that exist at European and national level are presented in the full report 
showing, among other aspects of the landscape, the legal frameworks governing compliance with sustainability.

\subsection{Setting the basis for sustainability}\label{sec:basis}

\begin{figure}[htb]
    \centering
    \includegraphics[width=0.7\textwidth]{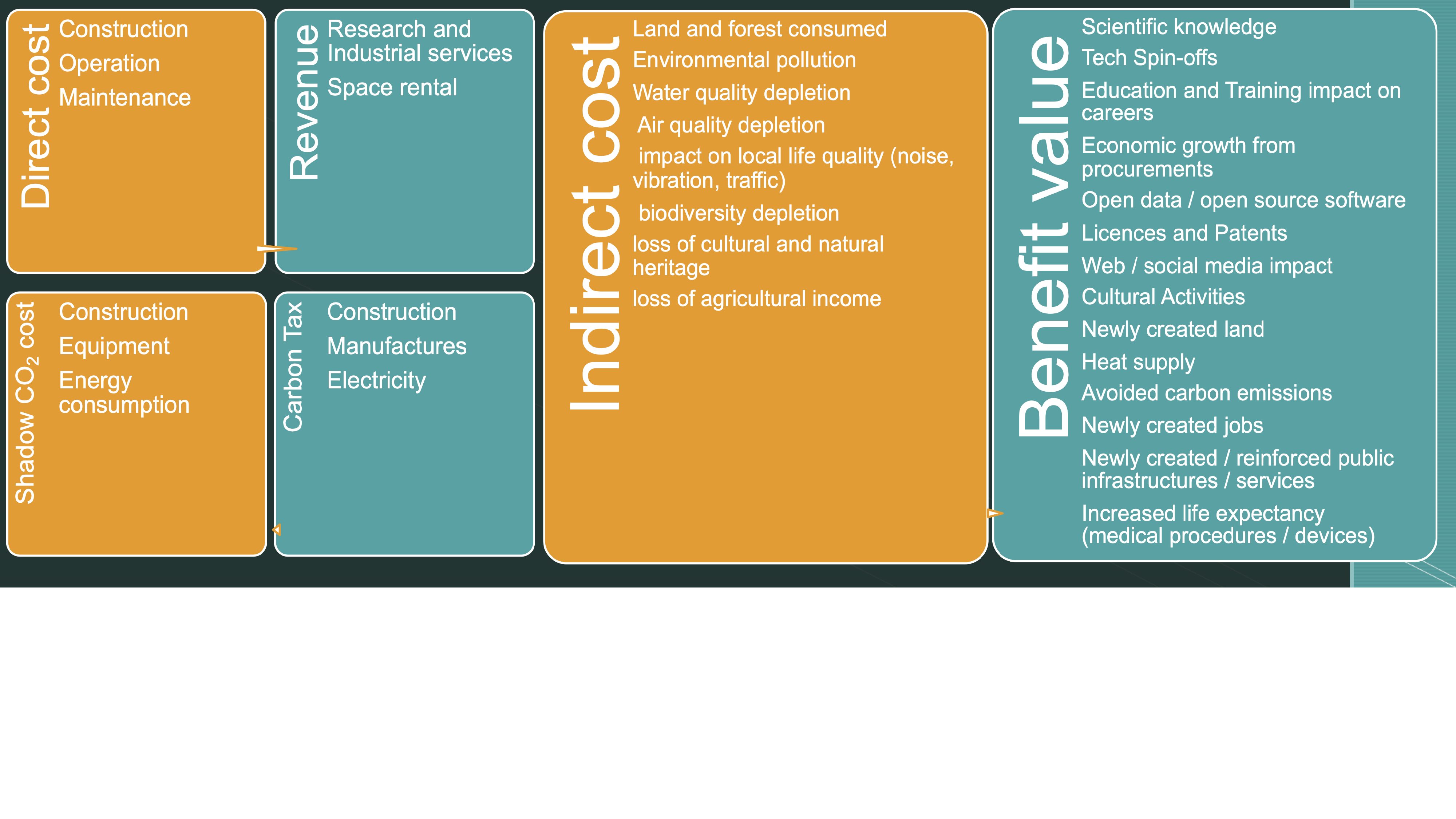}
    \vspace{-20mm}
    \caption{List of elements evaluated in a comprehensive cost-benefit analysis of research infrastructures.}
    \label{fig:CBAelements}
\end{figure}

A comprehensive sustainability assessment of particle accelerator-based research infrastructures -- as recommended by the Organization for Economic Co-operation and Development (OECD)~\cite{pearce2006cost, oecd2018cost}, 
the French government~\cite{quinet2014evaluation}, 
the UK government~\cite{UK_greenbook:2022}, 
and the European Commission~\cite{sartori2014guide} 
-- includes the evaluation of all costs and benefits of the project as presented in the example of \autoref{fig:CBAelements}.
Comprehensive sustainability assessments based on quantitative Cost-Benefit Analysis (CBA)~\cite{OMAHONY2021106587} 
integrate economic, social, environmental benefits, and total costs including indirect costs derived from environmental impact. 
Greenhouse Gas (GHG) emissions are identified and quantified in tonnes of carbon equivalent (t\ce{CO2}e) for each year of the observation period. In addition, the project-relevant avoided emissions are estimated, considering all mitigation measures adopted. Then, the t\ce{CO2}e are converted on a year-by-year basis in monetary terms using the \emph{shadow cost of carbon}~\cite[Sect.4]{EIBCBAGuide2023}, and quantified emissions are discounted using project-specific \emph{social discount rates} (SDR)~\cite[Sect.10]{EIBCBAGuide2023}. 
Finally discounted values are summed up, yielding a total ``cost'' of the carbon footprint of the project.
Considering that in several countries a carbon tax exists on different products, following the recommendations
of the World Bank~\cite{World_Bank_carbon_shadow_price}, the carbon tax paid on the consumed material and products is subtracted 
from the total monetary cost of carbon emissions  
to avoid double counting. 
The process is described in the 
``European Investment Bank Project Carbon Footprint Methodologies''~\cite{EIB_carbon_footprint}. 
This document recommends the use of the conversion rates of GHGs established by the Intergovernmental Panel on Climate Change (IPCC), regularly updated and available online~\cite{IPCC_GWPs}. 
Many different carbon shadow cost approaches exist~\cite{oeko_institut_2023}. 
To our best knowledge, the shadow cost of carbon established by the European Investment Bank (EIB) is the one most 
widely recognised and used in project appraisals. It is the recommended source to be used for converting t\ce{CO2}e to monetary cost~\cite{EIB_economic_appraisal}. 
While  costs, including negative externalities, are typically well defined in the existing guidelines, benefits and positive externalities are not fully captured. They are project-specific and proper evaluation of all externalities deserves further development. Relevant topics to consider and study are presented in section \ref{sec:SEse}. 

The 17 high-level development objectives of the UN Sustainability Development Goals (SDGs) illustrate the breadth of the economic, social, and environmental factors which can be influenced by the project. Organisations that develop, construct and operate particle accelerator-based research infrastructures are encouraged to identify and document quantified positive and negative impacts, considering and leaning on the UN Sustainability Goals as guiding framework~\cite{un-sdg-com}.

\emph{Environment and sustainability performance reporting.} 
Different schemes exist for reporting sustainability and environmental performance of organisations, projects and programs. 
Two prominent sustainability reporting schemes that are used worldwide are the Global Reporting Initiative (GRI), and the European Sustainability Reporting Standards (ESRS). Another simplified scheme widely used, albeit limited to environmental aspects, is the European Union Eco-Management and Audit Scheme (EMAS).

GRI~\cite{GRI} is an international independent standards organisation that aims at understanding and communicating a wide variety of sustainability-related impacts of organisations on issues such as climate change, human rights, and corruption. GRI's voluntary sustainability reporting framework has been adopted by multinational organisations, governments, small and medium-sized enterprises (SMEs), non-governmental organisations (NGOs), and industry groups. Over \num{10000} organisations from more than \num{100} countries use GRI, including CERN.
GRI started to integrate and align with the newly available European Union Corporate Sustainability Reporting Directive in 2021~\cite{EU_Corp_Sus_Reporting}. The European Sustainability Reporting Standards 
came into force on January 2024 and are rapidly evolving. EMAS is a tool to evaluate, report and improve the environmental performance of organizations and institutions. The regulation focuses on a standardised environmental statement, which can be certified by an independent environmental verifier. EMAS standardises a set of key performance indicators (KPIs) for the environment shown in \autoref{EIB:SCC}  that permits applying it to a project. 
\begin{table}[htb]
\vspace{-5mm}
\def\tw{\linewidth-3\tabcolsep}
\footnotesize
\centering
\caption{\small{EMAS key performance indicators for project-oriented environmental reporting.}}
\begin{tabular}{lp{0.5\textwidth}p{0.2\textwidth}}
\hline
\textbf{KPI} & \textbf{Description} & \textbf{Units} \\
\hline
Energy efficiency & Total annual energy consumption & MWh or GJ \\
Energy use & Percentage of energy from renewable sources used for electricity and heating & Percent \\
Material efficiency & Annual mass flow of different materials used & Tonnes per relevant material \\
Water & Total annual water consumption & Cubic metres \\
Waste & Total annual generation of waste broken down per type & Tonnes \\
Hazardous waste & Total annual generation of hazardous waste broken down per type & Tonnes \\
Biodiversity & Land surface used, consumed or constructed & Square metres \\
Emissions & Total amount of greenhouse gas emissions & Tonnes of \ce{CO2} equivalent \\
Air quality & Total annual emissions into air broken down by type & kg or tonnes \\
\hline
\end{tabular}
\label{EIB:SCC}
\end{table}
\vspace{-3mm}
\subsection{Socio-economic sustainability enablers}
\label{sec:SEse}
Socio-economic impact pathways are a valuable inventory of sustainability enablers. The EU-funded project \emph{RI-PATHS} has developed a toolkit~\cite{impact_toolkit} federating research infrastructures across various domains, including particle accelerator facilities such as ALBA, CERN and DESY.  
This section outlines how such pathways can directly lead to benefits and positive externalities with specific examples. 
Future projects are highly encouraged to carry out a comprehensive charting of the impact pathway potentials  
to develop sustainability-enabling plans.

\emph{Fundamental Physics Knowledge.} 
Science, providing  formalised knowledge that is rationally explicable and tested against reality, logic, and the scrutiny of peers, is a global public good~\cite{CouncilScience:2021}. In our transforming society, scientific knowledge is becoming the major factor in economic development, gradually replacing capital, land and labour. 
For particle physics, the opinion that scientific outcomes advance the society is shared by a large fraction of people, even if the asset is not directly used. 
Suitable approaches, such as the contingent valuation method also referred to as `stated preference', have been developed in economics to capture public perception of the value of science projects by estimating people willingness to financially participate (WTP).
These kind of surveys have been developed to analyse the value of outdoor recreational spaces~\cite{davis1963value}, for assessing levels for environmental protection~\cite{NOAA:1993}, and environmental incident mitigation measures~\cite{ExxonValdez:2003} as well as the investments to protect natural and cultural heritage. The methodology has been transferred to estimating the value of science, first for the LHC project~\cite{RePEc:mil:wpdepa:2016-03}, then for the HL-LHC project~\cite{CERN_Courier:2018} and for the FCC project~\cite{GIFFONI2023104627}. The approach has recently been adopted by ACTRIS, an atmosphere research infrastructure~\cite{Actris:2023}. 

\emph{Sustainability through education and training.} 
Particle accelerator and experimental physics research infrastructures offer the possibility to engage people at all education and training levels. 
If people have the opportunity to actively participate in the design, construction and operation of such projects, they enjoy benefits that translate directly into a \SI{2} to \SI{10}{\percent} lifetime salary premium~\cite{csil_2024_10653396, Camporesi_2017} compared to their peers that are enrolled in conventional training programs.

\emph{Sustainability through engagement of international research communities.} 
Engaging a large and continuously growing community of scientists and engineers in long-term collaborative research projects supports the sustainability of a scientific program. This partnership distributes personnel costs across multiple contributing institutions and countries, ensures a steady output of scientific products~\cite{csil_2024_13920183, MORRETTA2022121730} -- including Open Access publications -- and facilitates effective knowledge transfer into educational curricula through publications and direct training across generations. Finally, the concept of Open Innovation that engages a large number of knowledge domains ensures that the challenges of the science project will eventually spillover to domains that are immediately relevant to society~\cite{Gutleber2025}.

\emph{Sustainability through industrial spillovers.} 
Industrial spillovers from science projects generate direct benefits for industry and society~\cite{Griniece:2020, Castelnovo:2018, Florio:2016}. This impact pathway is more effective when science projects co-construct their instruments and infrastructures with industrial partners~\cite{Florio:2818, Autio:2004, Nordberg:2003}, and  companies work closely with the research project to develop technologically-intensive solutions delivering non-standard services~\cite{Sirtori:2019}. The approach is more laborious than a conventional client/supplier relationship since it requires exchanges of ideas and knowledge, the development of integrated processes, the mutual adaptation of working methodologies, and risk sharing. This process, however, leads to products and services that the industrial partners can leverage in other markets with earning multipliers above three, lasting  five to eight years. Industrial spillover can also lead to positive environmental externalities. The challenge to reduce the environmental impact of a new particle accelerator creates an opportunity for industrial innovations, developing, for instance, low-carbon concrete production process or improving construction techniques that make use of natural materials as wood and compressed earth. 
Environmental benefits can also be generated in the area of technical infrastructures that are developed with industrial partners. They range from high-efficiency refrigeration systems, to low-loss DC-based electrical systems, short- and medium-term energy storage,  waste heat buffering and supply, high-speed power control management, and machine learning infrastructure.

\emph{Sustainability through open information and computing technologies.} 
The development of Information and Communication Technology (ICT) and the generation of widely available data not limited to scientific results (e.g. engineering test data, operation monitoring, system tests) are outputs of particle accelerator-based research infrastructures~\cite{FLORIO201638}. Putting in place a global data sharing and processing infrastructure has already led to the creation of numerous, openly accessible software packages, platforms and online services, which are also used outside high-energy physics. They range from scalable data storage and distribution middleware to data management and workflow systems. Open-source software has been developed with value that extends beyond scientific collaborations. Examples include long-term data preservation, innovative Cloud computing services, meeting and event management software, particle-matter interaction modeling and analysis, and electronic library or information access software.  

\emph{Sustainability through cultural goods.} 
Creating public interest in science 
is part of the mission of the RIs.
Projects and organisations can develop a broad variety of activities, including 
permanent and travelling exhibitions, open days, guided tours, engagement with schools and teachers in joint workshops, citizen science projects, web sites, social media, engagement with video bloggers, on-line and TV documentaries, art internships, common art projects, feature movies, books, science fairs, presence in radio and TV shows and much more. 

\emph{Sustainability through positive environmental externalities.} 
Any future particle accelerator-based project has the potential to create positive environmental externalities that can compensate the residual, unavoidable and not further reducible negative environmental effects. Some examples are presented in \autoref{sec:mitigation}.
International and national legislation defines boundary conditions for the avoidance, reduction and compensation approach. The recently adopted update of the law for research and innovation in Switzerland for instance explicitly requires the consideration of the national and regional climate protection plans and energy-related aspects~\cite{fedlex:24.029}. Other countries, such as France, have already encoded the fight against climate change, resource and biodiversity protection, circular economy and sustainable territorial development in the environmental protection laws (L110-1 of~\cite{CodeEnvironnementFrance} and~\cite{Loi_2021_1104}) that govern the authorisation of new projects~\cite{Loi_2021_1104}. If properly planned, evaluated and monitored, collective compensation can even lead to a net positive effect~\cite{Cerema:2018}. However, some countries, such as Switzerland, require 1:1 compensation in the same region~\cite{BAFU:2022, Fedlex:451}. 

\emph{Innovation and R\&D.} 
New accelerators and particle detectors 
are projects spanning several decades while using cutting-edge technology. Both tend to be based on a technology that does not exist at the time of design. 
Novel technologies are designed and developed in-house or in partner institutions, in conjunction with external companies. The development can take the form of joint design, or procurement of technologies to be developed by the company.  
In both cases the industries profit from technological knowledge transfer, which may lead to technological breakthrough, patents or new business opportunities~\cite{bastianin2019technologicallearninginnovationgestation,Sirtori:2670056}. 

\emph{International collaboration.} 
Large RIs require international collaboration to be built and operated.
They are accountable and inclusive enterprises with a unique scientific mission and thus encourage countries to create common strategies at a ministerial level to pursue joint objectives. Developing a coordination network among different geopolitical actors entails agreements not limited to science and technology. 

\subsection{Life Cycle Assessment}
\label{sec:lca}

Future accelerator projects are encouraged to perform Life Cycle Assessments (LCA) at each stage  to 
understand the most critical
elements in the design, to provide input for technological choices reducing the environmental impact, and to fulfill
requirements for project submission.
This section, detailed in the full report, introduces the evaluation of environmental impact factors through an LCA, and presents some specific recommendations.
LCA is a methodological approach to quantify the environmental impact of a system, product or organisation.
The 14040 suite of ISO standards~\cite{iso14040:2006}, defines how such an assessment is conducted. 
It consists of four stages:
 i) definition of goal and scope, ii) life cycle inventory (LCI), iii) life cycle impact assessment (LCIA), and
    iv) life cycle interpretation.
Notably, an LCA is an iterative process, which means that new findings can be implemented in previous stages of the LCA and then run through again. 

\begin{figure}[h]
    \centering
    \includegraphics[width=0.9\textwidth]{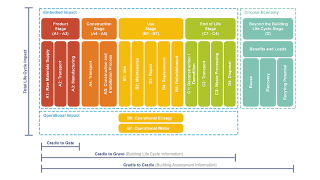}
    \vspace{-5mm}
    \caption{\small{Life cycle stages according to EN~15978~\cite{en15978:2011}, from D. Overbey~\cite{Overbey:2021a}, by author's permission}}
    \label{fig:lca-stages}
\end{figure}

The motivation and goals for a full, partial or simplified LCA depend on the phase of an accelerator project, ranging from an early concept phase or the engineering design phase to the construction or operation phase. Many existing and planned accelerator laboratories worldwide are already committed to the goal of conducting their scientific mission in a responsible and sustainable manner~\cite{CERN:2023a, GreenILC, CCC, ISIS-II, ESS}. Performing LCAs for existing or proposed accelerator projects contributes to this goal as it provides a solid basis to reduce environmental impact through optimising the design and operation of the RIs. In the planning phase, LCAs enable comparison and selection of design alternatives.
During this phase, the lever arm for carbon and environmental impact reduction is at its maximum, as design decisions are at their most flexible stage and have the added benefit of not being as expensive to make or change as they would have been in the later stages of construction or operation~\cite{pas2080:2023}.
In the engineering phase, LCAs support decision making on a more detailed engineering level, and are part of the preparation for project approval, where nowadays a quantification and justification of the carbon emissions is a legal requirement. The goals have an impact on the methodology and metrics adopted, including the scope covered and the impact categories studied. The scope definition of an LCA addresses three aspects: the functional unit, the system boundary, and the methodology.
For an accelerator project, the functional unit would be the accelerator, including detectors and supporting infrastructure.  
System boundaries are defined in relation to a life cycle model.
For instance, the European standard EN~15978~\cite{en15978:2011} defines a life cycle model (\autoref{fig:lca-stages}) in a manner that is also suitable to other large scale projects such as accelerators, with four main phases, i.e. product stage, construction stage, usage stage, and end of life.
A related, though slightly different way to categorise LCA scope is defined in the Greenhouse Gas Protocol~\cite[section 4]{Ranganathan:2004a}. The distinction is widely used when reporting emissions during the operation stage, in particular of an organisation or company.

\emph{Scope 1} refers to emissions from sources under the direct control of an organisation (e.g. from combustion of fossil fuels for transport or heating, or emissions from leaky gas systems)

\emph{Scope 2} covers emissions from bought-in energy - this almost exclusively refers to grid electricity consumption

\emph{Scope 3} covers all other indirect emissions, from sources outside of the RI direct control, such as embodied emissions in manufactured goods, commuting, business travel, data transfer and processing.

\noindent In the context of an accelerator project, Scope 2 and Scope 3 are the largest contributors to emissions: Scope 2 is dominated by the electricity needs during operation, while Scope 3 covers the emissions caused by the raw material production  and the manufacturing of accelerator components. 

Defining the methodology entails a choice of the impact categories to be evaluated, and the specifics of how the impact is calculated, for instance whether global averages or local impact factors are used for electricity or raw materials.  Emissions of substances influence the environment in different ways: greenhouse gas emissions change the energy imbalance of the earth, causing global warming; emission of fluorocarbons causes ozone layer depletion; emission of phosphate into water causes eutrophication. 
These different pathways of environmental impact are referred to as impact categories.   
The impact of emissions on climate change, quantified as global warming potential (GWP), is by far the most prominent in public debate and legislation. 
The overall warming caused by a specific substance depends on its radiation absorption profile and on how long it remains in the atmosphere, which in turn depend on the location of the emission. 
To make emissions of different substances comparable, weighting factors have been evaluated to obtain the impact of a specific substance with respect to \ce{CO2}, used as benchmark. 
The GWP over a 100 year period ($\mathrm{GWP}_{100}$) is given in relation to the emission of \ce{CO2} and thus measured in kilograms of equivalent \ce{CO2} emission, \unit{\kg \ce{CO2}e }. 
A number of methods and standards exist that specify how GWP should be evaluated and reported, among them: the Greenhouse Gas Protocol GHP~\cite{Ranganathan:2004a}, European standards EN~15804~\cite{en15804:2022}, and British Standards Institution (BSI) Publicly Available Specification PAS 2080~\cite{pas2080:2023}. 
These methods are all based on the approach defined by the Intergovernmental Panel on Climate Change (IPCC)~\cite[Tab. 8.A.1]{IPCC:AR5:WG1:Ch8}.

\noindent Beyond GWP, the so-called midpoint impact categories provide a quantitative way to compare emissions of different substances with respect to the same impact pathway. 
The weighting factors for midpoint impact categories are typically well defined and can be determined to reasonable accuracy. 
Midpoint categories, however, do not answer the question whether it is preferable to reduce emissions that cause harm via one impact category (say, global warming) at the expense of harm via a different pathway (say, eutrophication). 
For instance, replacing resistive electromagnets by permanent magnets poses such a dilemma: the permanent magnet option reduces the GWP, even when the higher carbon intensity of the magnet material is taken into account~\cite{shepherd:2023a}.
However, the use of materials with a high content of rare earth elements such as samarium or neodymium leads to increased impacts such as ecotoxicity, as well as social effects that are not classified as environmental impact, such as the impact of mining on local communities. 
Endpoint categories address this point by attempting to quantify the consequences for humans (measured in disability adjusted life years lost in the human population), damage to ecosystems (measured in number of species lost) and damage to resource availability. 
The endpoint categories are calculated from the midpoint categories by the application of weighting factors.
A number of methodologies exist that define sets of impact categories in addition to GWP, which is present in all methodologies.
A survey of LCA practitioners~\cite{Rybaczewska-Bazejowska:2024aa, Wahl:2018a} has shown that of these rather similar LCA approaches, ReCiPe 2016~\cite{Huijbregts:2017a,Huijbregts:2020a} is the most widely used.
An important part of the methodology is the choice whether specific or generic impact factors are used in the evaluation.
Impact factors can vary wildly, depending on the actual production process used, the country of origin of raw materials and location of a processing site, and on the time when a specific product has been or will be produced.
The choice of impact factors depends on the project stage and the goal of the LCA.
In an early phase of the project 
globally averaged values will be appropriate for most of the  materials evaluated; however, materials for civil construction and electricity will typically be drawn from the host country and thus local impact factors should be used. 
For projects that are envisaged to be realised far in the future, projections of impact factor reductions may also be taken into account, however, reliable projections are hard to find and will typically be limited to GWP reduction.
Usage of specific products, such as \ce{CO2}-reduced concrete, is difficult to take into account when purchasing policies are not known; such reductions can be identified as possible savings, or they might be considered if there exists a plan or even pledge to utilise such materials, possibly at a higher cost. 
LCA evaluations in the context of project appraisal will be based on more specific information about available suppliers and timelines, and may come with an obligation from the authorities to adhere to the values given. 
In such a case, actual product declarations (EPDs) may be available from potential suppliers.

The number of LCA publications has grown exponentially in recent years. For commonly used materials and components, such as simple resistors or capacitors, comprehensive sets of results are available. However, for accelerator-specific components and materials such as high-purity niobium or magnetic materials such as samarium-cobalt, publications are rare and 
calculated values for these materials vary widely. In order to ensure comparability, it is necessary to use common standards and also to agree on the methodology.
We summarize what is relevant for future accelerator projects to accurately, efficiently and consistently perform LCAs at various stages of the design and accelerator construction. 

 - At the early conception phase, efforts should be spent on performing a simplified LCA that uses generic data and focuses only on essential environmental aspects ( e.g. GWP [ \unit{\kg \ce{CO2}e} ] ), setting up the basis for LCAs at later design stages. The environmental consequences of the entire facility, including construction, operation and decommissioning impacts should be evaluated within the LCA. 
Extrapolations in time will be necessary and care should be taken to accurately estimate and take into account various scenarios (e.g. electricity supply decarbonisation policies), assumptions and uncertainties. 
In the optioneering stage, simplified LCAs should be performed for considered design options, to identify the differences of environmental impact. 
In the design stage, simplified LCAs should be performed for major components, and for accelerator design options to identify environmental impact drivers.
LCAs that have already been performed and are available for use e.g., sub-systems and components, should be incorporated into these LCAs. 
Any assumptions or alterations made e.g., the scaling of the LCA results of a magnet, should be documented.

 - Use the performed LCAs to identify areas of potential for reduction in environmental impact and take action to realise these reductions where feasible.

 - Use a common, widely-used and accessible Life Cycle Impact Assessment method (LCIA) with a range of indicators such as the ReCiPe 2016 midpoint (H) or the ILCD method. 

- Conduct dedicated LCAs of components widely used in accelerator projects such as superconducting cavities, superconducting magnets, or NEG coated vacuum chambers.  Results should be published to facilitate future evaluations of those facilities adopting the same or similar components. 

- Availability of common benchmark data will enable the comparison of LCA results from different projects and components. Many materials used in the accelerator construction have specifications that require special production processes, giving additional environmental impacts.
For some of these materials, for instance niobium, commercially available LCA databases contain either insufficient or inapplicable data for accelerator use cases. 
A common material reference database for accelerator projects should be compiled.

- The lifetime and downtime of the accelerator components and materials should be considered to estimate the full environmental impact of their uses at an accelerator.

\section{Environmental Impacts of Large Facilities}  
\label{sec:GHG_op}
This chapter describes the application of LCA  
 to the design, procurement and operation of accelerator facilities and includes some indications 
to mitigation strategies, further discussed in \autoref{sec:mitigation}.

\subsection{Accelerator construction and civil engineering works}

The construction phase (stages A1 to A5 of the LCIA process, as sketched in \autoref{fig:lca-stages}), covers the environmental impact up to the start of accelerator operation, considering all processes necessary to build civil engineering infrastructure and components, from raw material supply (stage A1), transport (A2) to manufacturing (A3), as well as the assembly, from transport (A4) to installation at the site (A5).

The laboratories that build and operate large accelerator-based infrastructures have experience in defining and working with complete and consistent \emph{product} and \emph{work breakdown structures} (PBS and WBS), that constitute a solid basis for gathering LCI data. Parts or subsystems that are reused across the project can be selected and analysed as separate functional units such that data gathering and analysis is done only once in the LCI process. LCI entails more than a tally of raw material masses. Depending on the subsystems and components under consideration, resources and emissions from fabrication and transport are relevant, as well as the amount of scrap material. Dedicated LCA studies of typical accelerator components such as magnets, power supplies, vacuum chambers or RF accelerating cavities are required to provide more precise input to project assessment. 
We recommend making estimates of quantities of materials required for the construction, as well as evaluation of approximate power consumption levels as early as possible, refining these estimates as the design progresses. 
\autoref{tab:project_phases} shows the typical phases of a large project, with the expected levels of detail for the environmental impact assessment at each phase, i.e. \emph{project definition phase}, \emph{conceptual design phase}, \emph{technical design phase}, \emph{authorisation-process phase}.
\begin{table}[h!]
\caption{
\small{Expected level of detail of environmental impact assessments 
for each phase of a large RI project.}}
\footnotesize
\centering
\begin{adjustbox}{width=\textwidth}
\begin{tabular}{p{2cm} lcccc p{3cm} l p{6cm}}
\hline
\textbf{Phase} & \textbf{Inventory} & \multicolumn{4}{c}{\textbf{Include in impact assessment}} & \textbf{Impact categories} & \textbf{Scope for} & \textbf{Goals of project} \\
& \textbf{detail level} & \textbf{Examine} & \textbf{Mass of} & \textbf{Factory} & \textbf{Transport} & & \textbf{mitigation} & \textbf{sustainability team} \\
& & \textbf{options} & \textbf{materials} & \textbf{processing} &  &  & & \\
\hline
Definition & Low & \tick & \tick & \cross & \cross & GWP only & High & Find likely largest-contributing areas and potential mitigation strategies \\
Conceptual design & Medium & \tick & \tick & \cross & \cross & GWP and others where data is readily available & High & Evaluate technology choices, including environmental impact as a factor \\
Technical design & Medium & \cross & \tick & \tick & \cross & All: decide on assessment methodology at this point & Medium & Fine-tune details of design; look for further reductions (e.g. alternative sourcing for raw materials) \\
Authorisation process & High & \cross & \tick & \tick & \tick & All & Low & Evaluate suppliers, including environmental impact; 
report 
to funding bodies \\
\hline
\end{tabular}
\end{adjustbox}
\label{tab:project_phases}
\end{table}

\noindent Civil engineering works give a substantial contribution to the environmental impact of a project. Excavation materials and concrete shielding constitute core subjects of the sustainability assessment. 
Reuse and recovery of the inert materials extracted from excavation of underground structures are two ways of preserving natural resources and facilitating the achievement of regulatory obligations.
A non-exhaustive list of potential reuse include i) use of specific excavated materials (e.g. limestone) in making cements and in stabilising constructions; ii) reuse of the excavated materials in backfilling quarries and mines; iii) development of new construction materials containing some specific constituents of the excavated materials (for example materials used in sandwich construction and tunnel stabilisation and insulating foams) for use within the project, where technically appropriate, or outside the project, if there is a market for them; iv) use of excavated materials in innovative pathways, for example processing of it into fertile soil for use in the re-naturation of wasteland and quarries or agricultural and forestry applications. 

\subsection{Accelerator and detector operation}

The operation phase (stages B1 to B7 of the LCIA process, as shown in \autoref{fig:lca-stages}) covers the environmental impact of the operational lifetime of an accelerator project; it covers the impacts of all processes necessary to operate (B1) and maintain (B2) the accelerator and experiments, including repairs (B3), replacement of components (B4) and refurbishments (B5). 
Of particular interest are the impact of energy (B6) and water (B7) required for operation.

\noindent Assessments of future accelerator projects concentrate frequently on  electricity consumption, assuming that it will  constitute the dominant impact, at least in terms of green house gas emissions.
This is, however, only valid assuming that future RIs overcome the current situation, in which direct emissions from particle detector and detector cooling plants are the most abundant sources of GHG emissions.
Several gaseous detector technologies make use of gas mixtures with very high GWP impact which have been chosen to obtain 
optimal detector performance and avoid ageing effects. The gas mixtures most in use are the \ce{C2H2F4} (known as R134a, GWP of 1430) and \ce{SF6} (GWP of 23900) for the Resistive Plate Chambers, the \ce{C4F10} (GWP of 8860) for Cherenkov detectors and the \ce{CF4} (GWP of 7390) for wire chambers, Cherenkov detectors and micro pattern gaseous detectors (MPGDs). These gases are necessary to mitigate ageing phenomena and to contain charge development (thanks to their electronegative properties) or to improve time resolution.
In the specific case of CERN, the annual emission during LHC Run 2 was around \SI{180}{\kilo\tonne\ce{CO2}e} (in 2022), \SI{50}{\percent} of which came from particle detectors of the LHC experiments. 
The emissions were due for around \SI{20}{\percent} to specific cases where it was not possible to recirculate \SI{100}{\percent} of the gas due to detector constraints, and for about \SI{80}{\percent} to the presence of leaks at the detector level in the ATLAS and CMS RPC systems. These leaks are mainly due to the breakage of plastic pipes and connectors, which break due to built-in fragility and mechanical stress. The leaks are not accessible during LHC run periods and, in some cases, also during end-of-the-year technical stops. 
The strategy to fulfill the objective to reduce GHG direct emissions is  based on i) \emph{gas recirculation}, with plants that purify and send back output gases from the detectors; ii) \emph{gas recuperation}, with the gas mixture sent to a recuperation plant where gases are extracted, stored and re-used; iii) \emph{study of alternative gases}, suitable for particle detectors and with much lesser GWP impact.

Emissions produced in operating an accelerator arising from electricity production (i.e. scope 2 emissions) depend on the grid intensity of the country hosting the facility, which is a function of the energy generation mix within that country (coal, gas, nuclear, wind, solar) and is expected to vary with the year of operation. Examples of the levels of carbon emissions associated to the production of electrical energy in a few representative countries and years are shown in \autoref{ElectricityCarbonFootprint}. 
\begin{table}[ht]
\caption{\small{Examples of some selected official electrical energy carbon footprint sources.}}
\label{ElectricityCarbonFootprint}
\centering
\footnotesize
\begin{tabular}{lllcc}
\hline
\textbf{Country} & \textbf{Source} & \textbf{Energy} & \textbf{Carbon footprint} & \textbf{Year} \\
&&&[\unit{\kg \ce{CO2}e \per\mega\watt\hour}]& \\
\hline
France & Ademe Base Empreinte & Offshore wind & 15.6 & 2023 \\
France & Ademe Base Empreinte & Unqualified energy mix & 52.0 & 2022 \\
Germany & Umweltbundesamt (UBA) 
& Unqualified energy mix & 380.0 & 2023 \\
Italy & ISPRA & Unqualified energy mix & 257.2 & 2022 \\
Switzerland & BAFU/OFEV 
& Unqualified energy mix & 54.7 & 2018 \\
Switzerland & BAFU/OFEV & Renewable energy mix & 15.7 & 2018 \\
\hline
\end{tabular}

\end{table}
A proper energy management plan, as defined by ISO 50001 standards, is needed to implement a policy of prioritized actions to 1) avoid, 2) reduce, and 3) re-use energy.
Continual improvement on energy performance is the ultimate goal of an ISO 50001 Energy Management System (EnMS). It requires the development of a policy for the efficient use of the energy, i) looking closely at the specific operation of the accelerator-based RI, ii) identifying targets and objectives to be reached, iii) implementing procedures and equipments to gather data on energy consumption, and iv) planning regular review processes to be used for reporting and improving the EnMS. 
The largest subsystems by power usage are i) the plants for air conditioning and water cooling, accounting for about \SI{35}{\percent} of the electrical power requirements, and  ii) the RF systems, where RF power is generated, transported through waveguides and delivered to the cavities which accelerate particle beams, and each stage is affected by efficiency losses.
In case of superconducting RF, the cavities (typically composed of solid niobium) must be cooled to \SI{1.8}{\kelvin} to \SI{2}{\kelvin}. 
Since the operation of the RF depends only on current flowing on the inner surface of the cavity, it is possible to use an alternative construction, with a thin film of superconducting material deposited on a solid copper substrate. Copper provides mechanical rigidity whilst the thin film of niobium (or other superconductor) provides RF properties. Thin-film cavities can operate at a higher temperature than bulk cavities (\SI{4}{\kelvin} rather than \SI{1.8}{\kelvin}), which drastically 
reduces the amount of power required for the cryogenic system, and reduce the amount of niobium needed in the construction. 

Environmental impacts from maintenance (B2), repair (B3), and replacement (B4) have been rarely considered in accelerator LCAs. 
Expenditures for the procurement of spare parts for those processes is an important contribution to the operating costs of an accelerator, and empirical data exists on the typical lifetime of key components such as klystrons that have to be replaced on a regular basis. 
This data can be used, together with other materials, such as helium or liquid nitrogen which are regularly replenished, 
to quantify footprint coming from accelerator maintenance, repair and replacements. 
Front-end electronics and computers are needed to operate large particle detectors of the experiments at the accelerator-based RIs. 
The production of silicon chips uses a lot of resources, as discussed for instance in ref.~\cite{WANG202347}. 
Typically, server-chips are "refreshed" every 18 to 24 months so that replacements, together with the consumption of electrical energy to switch and then to cool the transistors,
contribute to the footprint of the operation phase. 
A dedicated section of the full report presents the landscape of  detector-processing computing systems, and the open technological options for data processing, cooling and heat re-use.

\subsection{Decommissioning}

The decommissioning phase of an accelerator (stages C1 to C4 of the LCIA process, as sketched in \autoref{fig:lca-stages})  covers the environmental impacts after the end of operation until all components and materials are properly disposed and consists of: demolition stage (C1), transport of materials (C2), processing of waste (C3), and final disposal (C4).

\noindent Evaluation of the impact of the decommissioning is subject to a number of rather arbitrary assumptions. 
First, accelerators are very long-lived, and the end of life of its components can vary considerably.
For example, CERN's proton synchrotron, originally commissioned in 1959, today, 65 years later, is still an integral part of the LHC injection chain.
Other accelerators, for example HERA at DESY, have been decommissioned, but still await final demolition and disposal of components, many of which are still re-usable.
Some accelerators, such as LEP at CERN, have instead been demolished completely. A report~\cite{international2020iaea} commissioned by the International Atomic Energy Agency (IAEA) has collected valuable information on this process. 
Recently, the procedures to re-classify radioactive materials from accelerators at CERN have been published~\cite{Svihrova:2024tzy}.

An issue particular to the disposal phase of accelerator projects is the treatment of activated material.
The disposal life cycle of material that has been subjected to radiation depends on national regulations. 
Therefore, defining material categories such as ``stainless steel, induced activity below a threshold in  \unit{\micro\sievert\per\hour}'' or ``stainless steel, total irradiation below a threshold in \unit{\mega\gray}'' that correspond to a specific disposal path, is difficult and is not implemented yet in an LCA.
Predicting the amount of radiation materials requires dedicated, complex simulations of radiation sources, material composition and operating scenarios, which is typically done only for certain parts of the accelerator, and typically only in the engineering design phase of a project. 
Most of the radioactive waste that is generated in particle accelerators can be classified as waste with very-low-level (VLL)~\cite{international2020iaea} activity, which can also be candidate for clearance from regulatory control (also called free-release) in those countries where this procedure is established. 
There is an environmental and economic benefit in performing the release from regulatory control whenever this procedure can be applied. 
Instead of being stored for decades in a repository, released material is recycled in the metal industry, thus reducing raw material footprint. 
Targets, beam-dumps and any other accelerator components that are directly hit by the beam can reach low- to intermediate-levels of activity. The disposal of activated waste towards final repositories requires accurate radiological characterisation to ensure that the activity they store falls below the activity limits for which the repository was designed.
The possible disposal pathways depend on the national regulations and in Europe 2006/117/Euroatom and 2011/70/Euroatom are the directives to be adopted by its member states. 

The early consideration of waste disposal pathways when designing an experiment or a facility not only contributes to the minimisation of  radioactive waste, but also simplifies the entire industrial disposal process. 
From an LCA perspective, \emph{minimisation of the production of radioactive material and waste} is extremely important and extensively discussed in the full report. 

\subsection{Data on Future Collider Projects}

A synthesis of data from sustainability assessment of the accelerator projects is requested in public debates, within research communities and science management panels.
Proposals of future accelerators all contain running scenarios identified by center-of-mass (CoM) energy, average luminosity, operation duty cycles,  integrated luminosity per year, number of interaction regions (IPs), and include upgrades of different kinds. 
Table \ref{tab:ghg-project} presents a list of key parameters relevant for GHG emissions in construction and operation phases of future collider projects, as discussed and agreed by the Sustainability WG. Concurrent reporting of methods, assumptions and simplifications adopted to evaluate carbon footprint is needed to clarify the boundaries of the assessment.   
Data from several projects of future circular (CEPC, FCCee, and LHeC) and linear (CLIC, ILC, LCF, and C$^3$) colliders are presented in the full report.
\begin{table}[htbp]
    \caption{\small{Key parameters and estimated GHG emissions of future colliders projects}}
 \label{tab:ghg-project}
    \footnotesize
    \centering
   \begin{tabular}{l|c|c|c|}
    \hline 
      \textbf{Project } & Project phase A & Project phase B& Project phase ...\\
 \hline
  CoM energy [\unit{\GeV}]  & & & \\
  Luminosity/ IP $[10^{34} \unit{\cm^{-2}}\unit{\s^{-1}}~] $ &  & & \\
  Number of IPs &  & & \\
  Operation time for physics/ yr  $[10^7 \unit{s}/ \unit{yr}]$ &  & & \\
  Integrated luminosity/ \unit{yr}  [1/\unit{fb}/ \unit{yr}]  &  & & \\
  Host country & 
           \multicolumn{3}{c|}{ } \\
   \hline 
  \multicolumn{4}{l|}{\textbf{GHG emissions from construction, stage A1-A5}} \\
  \hline 
  Tunnels, caverns, shafts [\unit{kt}~\ce{CO2}e] & & & \\
  Surface buildings [\unit{kt}~\ce{CO2}e] & & & \\
  Accelerator (coll.) [\unit{kt}~\ce{CO2}e] & & & \\
  Accelerator (inj.) [\unit{kt}~\ce{CO2}e] & & & \\
  Detectors [\unit{kt}~\ce{CO2}e] & & & \\
  \textbf{Total [kt~\ce{CO2}e]} & & & \\
  Collider tunnel length  [\unit{km}] & & & \\
  Collider tunnel diameter [\unit{m}] & & & \\
  Collider tunnel GHG / m  [\unit{t}~\ce{CO2}e/ \unit{m}] & & & \\
  Concrete GHG  [\unit{kg}~\ce{CO2}e/ \unit{kg}] & & & \\ 
  Accelerator  GHG / m  [\unit{t}~\ce{CO2}e/ \unit{m}] & & & \\
  \hline 
  \multicolumn{4}{l|}{\textbf{GHG emissions from operations}} \\
  \hline 
  Total power in operation [\unit{MW}]   &  & & \\
  Electricity cons. / yr  [\unit{TWh}/ \unit{yr}]  & & & \\
  Year of operation & & & \\
  Carbon intensity of electr. [\unit{g}~\ce{CO2}e/ \unit{kWh}] & & & \\
  Scope 2 emissions / yr [\unit{kt}~\ce{CO2}e] & & & \\
    \hline 
    \end{tabular}

\end{table}

\section{Mitigation and Compensation Measures}
\label{sec:mitigation}

Sustainability assessment also includes the evaluation of the adverse environmental effects avoided by implementing mitigation strategies and compensation measures.
The development of responsible energy management plans (introduced in section 3.2), the implementations of eco-designs aiming to reduce energy consumption, adoption of responsible procurement procedures, commitment to R\&D improving the energy efficiency of the accelerator, selection of better/greener materials for civil engineering works, heat recovering and redistribution, and nature-based interventions are all topics discussed in the full report.     

\emph{RI design with sustainability as a goal.} Early consideration of civil infrastructure designs aiming to optimize the tunnel structure and minimize material consumption, and that envisage the use of low carbon concrete, can significantly reduce the impact on the environment of the construction phase.
For example, studies of the LCA of the construction of CLIC and ILC, including  underground facilities, covering tunnels, caverns and access shafts~\cite{clic_ilc_lca_arup} suggest that embodied carbon emissions could be reduced by 40\% with a design and material selections that integrate carbon reduction
to technical specifications.  

\emph{Responsible procurement.} Shifting towards responsible procurement implies the balancing of the initial, purchasing cost with i) total lifecycle cost (including maintenance, operation, disposal); ii) environmental impact against a given set of environmental objectives (e.g. reduction of \ce{CO2} emissions, minimisaton of radiological impacts); iii) social responsibility (e.g., ensuring that no child labour is involved in the manufacturing at any moment, including the extraction of ores and raw materials). A responsible procurement policy is one of the components of the overall strategy of scientific institutions to pursue
sustainability. Adoption of a responsible procurement policy based on the international standard  ISO 20400:2017, or any equivalent standard established in different regions of the world, is recommended. 
The  standard ``provides guidance on how sustainability considerations are integrated at a strategic level within the procurement practices of an organization, to ensure that the intention, direction and key sustainability priorities of the organization are achieved." It does not contemplate any specific certification. 
Concerning the use of electricity, planning for a portfolio of supply contracts or energy power purchasing agreements (PPA)\cite{gutleber_2023_10023947} constitutes a strategic choice to foster the transition to renewables and reduce the GWP impact of the accelerator-based RIs. Pooling suppliers in a sufficiently large procurement action that is particularly energy intense and establishing a dedicated energy supply contract for such a procurement action, is an approach that has been successfully implemented already by different countries and industry consortia~\cite{crescenzi_2024_13166167}. Finally, the European Union Emission Trading System (EU ETS), the world's first and largest carbon market, is an instrument that can be leveraged to compensate the residual carbon footprint and help bring overall EU emissions down while generating revenues to finance green transition.

\emph{Heat recovery and supply.} Most of the energy used to operate the technical infrastructures and subsystems of a particle accelerator is eventually converted into heat.
Energy used to operate accelerator magnets, amplifying radiofrequency energy, absorbing synchrotron radiation, air conditioning systems, operating electronics and data processing centers is almost entirely converted into low grade heat, typically below \SI{45}{\celsius}. Only rarely temperatures above \SI{50}{\celsius} are reached, for instance with cooling cryogenic refrigeration systems and electrical transformers and substations. Considering the amounts of heat that particle-accelerator based research infrastructures generate there exists an interest to explore ways to recover that heat and convert it into a valuable resource. A comprehensive cost-benefit assessment is needed to understand if heat recovery is viable for a specific project and to develop the most suited heat recovery and supply scenario for that project. The LCA approach can be integrated in this analysis to better quantify negative externalities.

\emph{Investment in R\&D on green technologies.} Among all actions focused on minimising the environmental impact of future colliders, study and development of novel technologies is the most specific to the accelerator-based RIs.
Medium- and long-term challenging development plans are in place, aiming to improve the efficiency of the RF systems as mentioned in section 3.2, to design advanced permanent magnets, to use high-temperature superconductors (HTSs) in high-field magnets, or to understand/demonstrate the full potentiality of energy-recovery linacs (ERLs).
Large initiatives are in progress to coordinate and foster the programs carried out by national laboratories worldwide. 
Horizon Europe fundings support research programs such as iFAST (Innovation Fostering in Accelerator Science and Technology) and iSAS (Innovate for Sustainable Accelerator Systems) committed to the development of innovative solutions to contain the energy demand of future accelerators, EAJADE (Europe-America-Japan Accelerator Development and Exchange)  that promotes the exchange of accelerator scientists between European institutions and American and Japanese partners, and RF2.0 (Research Facility 2.0 towards a more energy-efficient and sustainable path).
The Magnet Development Program (MDP) has been established by DOE to coordinate the R\&D activities in US laboratories. An explicit  acknowledgement of their key role in sustainability is expressed in the 2023 P5 Report  
drawn by the \emph{particle physics prioritization panel} on inputs, as ref.\cite{ITF}, from the \emph{Snowmass'21} process, discussed in the full report. 
In China, extensive efforts are devoted to sustainability studies by the IHEP towards the Circular electron-positron Collider (CEPC)~\cite{CEPCTDRDec24}. They are focused on energy recovery plants and on the efficiency improvement of RF systems, summarized in the full report.

\emph{Nature-based Interventions for Carbon Removal.} `Nature-based interventions' --- also termed `Nature-based climate solutions' (NbCS) --- are ``conservation, restoration and improved management strategies (pathways) in natural and working ecosystems with the primary motivation to mitigate GHG emissions and remove \ce{CO2} from the atmosphere''\cite{Buma:2024}. 
They can address climate change in different ways: i) by decreasing GHG emissions related to deforestation and land use; ii) by capturing and storing \ce{CO2} from the atmosphere; iii) by enhancing the resilience of ecosystems\cite{Miles:2021}.

\clearpage
\printbibliography

\end{document}